\documentclass[final,number,5p,times,square,twocolumn, longbibliography]{elsarticle}

\usepackage{inputenc}
\usepackage{graphicx}
\usepackage{verbatim}
\usepackage{epsfig}
\usepackage{bm}
\usepackage{color}
\usepackage{xcolor}
\usepackage{float}
\usepackage{dcolumn}
\usepackage{multirow} 
\usepackage{physics}
\usepackage{braket}
\usepackage{comment}
\usepackage{soul}
\usepackage{subcaption}
\usepackage[colorlinks=true,
            linkcolor=blue,
            citecolor=blue,
            urlcolor=blue]{hyperref}

\usepackage{ulem}
\newcommand\redsout{\bgroup\markoverwith{\textcolor{red}{\rule[0.5ex]{2pt}{1.4pt}}}\ULon}

\renewcommand{\emph}[1]{\textit{#1}}

\newcommand{\nc}{\newcommand}

\newcommand{\beq}{\begin{equation}}
\newcommand{\eeq}{\end{equation}}
\nc{\bfx}{{\bf x}}
\nc{\bfy}{{\bf y}}
\nc{\bfz}{{\bf z}}
\nc{\bfxh}{{\bf \hat{x}}}
\nc{\bfyh}{{\bf \hat{y}}}
\nc{\bfzh}{{\bf \hat{z}}}
\nc{\bfj}{{\bf j}}
\nc{\bfr}{{\bf r}}
\nc{\bfR}{{\bf R}}
\nc{\bfk}{{\bf k}}
\nc{\bfq}{{\bf q}}
\nc{\bfp}{{\bf p}}
\nc{\bfv}{{\bf v}}
\nc{\bfs}{{\bf s}}
\nc{\bfA}{{\bf A}}
\nc{\bfJ}{{\bf J}}
\nc{\bfsg}{{\bm \sigma}}
\nc{\bfvh}{{\bf \hat{v}}}
\nc{\bfqh}{{\bf \hat{q}}}
\nc{\low}{\delta_{\rm Low}}

\newcommand{\avg}[1]{\langle #1 \rangle}

\journal{Physics Letters B}

\date{LA-UR-26-24528}

\begin{document}

\begin{frontmatter}



\title{Quantum Monte Carlo calculations of Zemach moments in $A\leq 9$ nuclei}

\author[first]{Garrett B.~King}
\ead{kingg@lanl.gov}
\affiliation[first]{organization={Theoretical Division, Los Alamos National Laboratory},
            city={Los Alamos},
            postcode={NM 87545},
            country={USA}}

\author[second,third]{Sonia Bacca}
\ead{s.bacca@uni-mainz.de}
\affiliation[second]{organization={Institut für Kernphysik and PRISMA$^+$ Cluster of Excellence, Johannes Gutenberg-Universität},
            city={Mainz},
            postcode={55128},
            country={Germany}}
\affiliation[third]{organization={Helmholtz-Institut Mainz, Johannes Gutenberg-Universität},
            city={Mainz},
            postcode={55099},
            country={Germany}}

\author[fourth]{Graham Chambers-Wall}
\ead{chambers-wall@wustl.edu}
\affiliation[fourth]{organization={Department of Physics, Washington University}, city={Saint Louis}, postcode={MO 63130}, country={USA}}

\author[fifth,sixth]{Alex Gnech}
\ead{agnech@odu.edu}
\affiliation[fifth]{organization={Department of Physics, Old Dominion University}, city={Norfolk}, postcode={VA 23529}, country={USA}}
\affiliation[sixth]{organization={Theory Center, Jefferson Lab}, city={Newport News}, postcode{VA 23610}, country={USA}}

\author[fourth,seventh]{Saori Pastore}
\ead{saori@wustl.edu}
\affiliation[seventh]{organization={McDonnell Center for the Space Sciences at Washington University}, city={St. Louis}, postcode={MO 63130}, country={USA}}

\author[fourth,seventh]{Maria Piarulli}
\ead{m.piarulli@wustl.edu}

\author[eigth]{R.~B.~Wiringa}
\ead{wiringa@anl.gov}
\affiliation[eigth]{organization={Physics Division, Argonne National Laboratory}, city={Argonne}, postcode={IL 60439}, country={USA}}

\begin{abstract}

Modern atomic spectroscopy has reached a level of precision at which nuclear-structure effects can no longer be neglected and must be quantified reliably. In particular, hyperfine splittings depend on the Zemach radius, which encodes the convolution of the nuclear charge and magnetization distributions. The third electric Zemach moment provides a related finite-size measure and enters the elastic two-photon-exchange contribution to the Lamb shift in muonic atoms. Here, we compute Zemach radii and other electromagnetic moments for light nuclei using quantum Monte Carlo techniques within modern \textit{ab initio} nuclear theory. Using Norfolk two- and three-body interactions derived within chiral effective field theory, we assess the model dependence and study the role of two-body currents.  For $^6$Li, we obtain a Zemach radius larger than that extracted from atomic measurements, consistent with recent calculations, confirming that the discrepancy is not an artifact of the nuclear model. For $^9$Be, our results agree with experiment; the discrepancy of previous phenomenological evaluations is traced to a model-dependent input for the magnetic radius. 

\end{abstract}



\end{frontmatter}




\section{Introduction}
\label{introduction}

High-precision laser spectroscopy provides stringent tests of the Standard Model and offers a sensitive probe of possible physics beyond it. Recent advances in optical spectroscopy of trapped atoms and ions have reached an unprecedented level of precision, with relative accuracies at the parts-per-quintillion level~\cite{Berengut:2025nxp}. This progress opens new opportunities in the search for physics beyond the Standard Model~\cite{Safronova:2018}, providing constraints complementary to those obtained from high-energy colliders, astrophysical observations, and cosmology. However, at this level of precision, the interpretation of atomic measurements requires equally accurate nuclear-structure input. Nuclear effects must be reliably quantified in order to disentangle potential signatures of new physics from Standard Model contributions associated with the finite size and internal structure of the nucleus.

Among atomic observables, hyperfine splittings and Lamb-shift transitions are particularly sensitive to nuclear-structure effects. Hyperfine splittings arise from the interaction between the nuclear magnetic moment and the electromagnetic fields generated by the orbiting lepton, and their high experimental precision makes them a powerful probe of nuclear charge and magnetization distributions. 
The leading nuclear-structure correction associated with the convolution of the nuclear charge and magnetization distributions is commonly expressed in terms of the Zemach radius~\cite{Zemach:1956}. Zemach higher-order moments of these distributions that enter as subleading corrections-- increasingly relevant as experimental precision improves-- also play an important role in the interpretation of muonic atom spectroscopy.  In particular, the third electric Zemach radius, also known as the Friar moment~\cite{Friar:1979}, contributes to the elastic part of the nuclear two-photon-exchange correction to the Lamb shift~\cite{Ji:review}. Since these corrections are strongly enhanced in muonic systems, reliable nuclear-structure input is required for the extraction of charge radii and for the interpretation of precision spectroscopy measurements. This is especially timely  in view of the experimental program of the QUARTET collaboration~\cite{Quartet}, which aims at high-precision X-ray spectroscopy of light muonic atoms.

A simple phenomenological approach models both the nuclear charge and
magnetization distributions using Gaussian functions, with parameters fixed to
reproduce the empirical charge and magnetic radii~\cite{Yerokhin:2008}, typically coming from electron scattering measurements.
This kind of estimate, hereafter referred to as the Gaussian model and denoted by $\avg{R_Z^{\rm pheno}}$, is typically adopted as the nuclear physics reference in atomic physics analysis~\cite{Qi:2020,Sun:2023}. However, it is approximate and may not be sufficient at the level of precision targeted by current and future experiments, motivating a more rigorous \textit{ab initio} derivation. 

From the measured hyperfine splitting, an effective Zemach radius $\avg{R_Z^{\rm eff}}$ can be extracted after subtracting  point-nucleus quantum electrodynamics effects, recoil, and other well-controlled contributions. This effective quantity singles out the pure nuclear-structure correction and, in the absence of significant inelastic effects, is expected to agree with the Zemach radius computed from the convolution of the nuclear charge and magnetization distributions.  
However, as shown in Table~\ref{tab:nucl.phys.val}, significant discrepancies $|\Delta_{\rm exp}|$ between $\avg{R_Z^{\rm eff}}$ and $\avg{R_Z^{\rm pheno}}$ are observed in light nuclei, with particularly large deviations seen in $^6$Li and $^9$Be. 
The $^6$Li case is particularly interesting: despite having a larger charge radius than $^7$Li, it exhibits a Zemach radius which is approximately $40\%$ smaller, in tension with expectations based upon the simple Gaussian model. 
Recently, Yang et al.~\cite{Yang:2025rcx} proposed that the $^6$Li anomaly  could be explained by inelastic nuclear polarizability effects, which are present in the effective Zemach radius $\avg{R_Z^{\rm eff}}$ but not accounted for in the purely elastic term calculated from density distributions.

\begin{table}[tbh]
\centering
{\small 
\setlength{\tabcolsep}{3pt} 
\renewcommand{\arraystretch}{0.9}
\begin{tabular}{ c c l c c } \hline \hline 
Nucleus & $\avg{R_Z^{\rm pheno}}$ [fm]~\cite{Yerokhin:2008} & $\avg{R_Z^{\rm eff}}$ [fm] & $|\Delta_{\rm exp}|$ [fm] &$\%$ diff.\\ \hline
$^6$Li & $3.71(16)$ & $2.44(2)$ \cite{Sun:2023} & $1.27$ & $52.0$\\ 
& &$2.40(16)$~\cite{Qi:2020} &$1.31$ & $54.6$ \\
& &$2.47(8)$~\cite{Qi:2020} &$1.24$ & $50.2$ \\
& &$2.30(3)$~\cite{Puchalski:2013} &$1.41$ & $61.3$ \\ \hline
$^7$Li & $3.42(6)$ & $3.33(7)$~\cite{Qi:2020} & $0.09$ &$2.7$\\ 
& &$3.38(3)$~\cite{Qi:2020} &$0.04$ &$1.2$ \\ 
& &$3.25(3)$~\cite{Puchalski:2013} &$0.13$ &$4.0$ \\ \hline
$^9$Be & $3.38(2)$ & $4.07(5)(2)$ \cite{Puchalski:2014} & $0.69$ &$17.0$ \\ 
& &$4.048(2)$~\cite{Dickopf:2024} &$0.67$ &$16.5$ \\ \hline
\end{tabular}}
\caption{Zemach radii $\avg{R_Z^{\rm pheno}}$ calculated within the Gaussian model of Ref.~\cite{Yerokhin:2008} compared with  values extracted from atomic spectroscopy data,  $\avg{R_Z^{\rm eff}}$.
Their absolute difference $|\Delta_{\rm exp}|= | \avg{R_Z^{\rm pheno}}-\avg{R_Z^{\rm eff}}|$  and associated percentage discrepancy is also provided. 
}\label{tab:nucl.phys.val} 
\end{table}

In this work, we determine the Zemach radii and third electric Zemach moments of light nuclei using \textit{ab initio} quantum Monte Carlo (QMC) methods. Starting from nuclear wave functions obtained with realistic nuclear interactions, we evaluate the corresponding charge and magnetization distributions and calculate the associated Zemach observables together with their theoretical uncertainties. This approach allows us to assess whether the discrepancies observed in phenomenological extractions, in particular in the $^6$Li Zemach-radius puzzle, may originate from simplified nuclear structure treatments or instead point to neglected effects. Our results provide nuclear-structure input relevant for the interpretation of high-precision hyperfine-splitting and Lamb-shift measurements in electronic and muonic atoms.

\section{Zemach radii}\label{sec:model}
The definition of the classical Zemach radius in coordinate space is 
\begin{equation}
\label{eq:Zemach_r}
    \avg{R_Z} = \int d^3 r \, d^3 r' \, \rho_{\rm ch}(\mathbf{r}) \, \rho_{\rm M}(\mathbf{r}') \, |\mathbf{r} - \mathbf{r}'| ,
\end{equation}
where $\rho_{C}$ and $\rho_{M}$ are the charge and magnetic density distribution, respectively. 
In momentum space, this becomes 
\begin{equation}
\label{eq:Zemach_q}
    \avg{R_Z} = -\frac{4}{\pi\mu} \int_0^{\infty} \frac{dq}{q^2}\left[ F_C(q^2)F_M(q^2) - 1 \right]\, ,
\end{equation}
where $F_C$ and $F_M$ are the charge and magnetic form factors, respectively, and $\mu$ is the magnetic moment of the nucleus. We will calculate $\avg{R_Z}$ by using Eq.~(\ref{eq:Zemach_q}), following the integration procedure described in Ref.~\cite{NevoDinur:2018hdo} to perform the integration.  At sufficiently small $q$, the form factors behave as
\begin{equation}
    F_C(q^2) \approx 1 - \frac{1}{6}\avg{R_E^2}\,q^2 + \frac{1}{120}\avg{R_E^4}\,q^4 + \ldots\, ,
\end{equation}
and 
\begin{equation}
    F_M(q^2) \approx \mu q \, \left(1 - \frac{1}{6}\avg{R_M^2}q^2 + \frac{1}{120}\avg{R_M^4}q^4\right) + \ldots\, ,
\end{equation}
where $\avg{R_x^n}$ is the $n^{\rm th}$ moment of the ``$x$" distribution of the nucleus, $x$ being either charge or magnetic. We will present results for $R_{E/M} = \sqrt{\avg{R_{E/M}^2}}$, as they are used to approximate the integration kernel for small $q$ following the procedure of Ref.~\cite{NevoDinur:2018hdo}.

Another quantity of interest is the third electric Zemach radius (or moment), which contributes
  to the elastic two-photon exchange processes in Lamb shift transitions for muonic atoms~\cite{Ji:review}. In momentum space, it is defined as
\begin{equation}
    \avg{R^3_E}_{(2)} = \frac{48}{\pi} \int_0^{\infty} \frac{dq}{q^4}\left[ F^2_C(q^2) - 1 + \frac{q^2\avg{R_E^2}}{3} \right]\, 
\end{equation}
and will also be investigated in this paper. 

\section{Theoretical approach}\label{sec:model}
We solve the nuclear many-body problem using two quantum Monte Carlo (QMC) methods: namely, variational Monte Carlo (VMC) and Green's function Monte Carlo (GFMC), reviewed in Refs.~\cite{Carlson:2014vla,Gandolfi:2020pbj}.
We start from Hamiltonians which include two- and three-nucleon forces 
derived in a chiral effective field theory ($\chi$EFT) approach retaining nucleons, pions, and $\Delta$ isobars as degrees of freedom.
The two-body interaction contains one-pion exchange, two-pion exchange, and short-range contact terms. The contact terms in the two-nucleon force encode the short-range physics that is integrated out in the $\chi$EFT approach, and are thus parameterized by low-energy constants (LECs) that are fit to data. The three-nucleon force contains long-range pion exchanges, as well as contact interactions with unknown LECs. 

We work with interactions that belong to the Norfolk family ~\cite{Piarulli:2016vel,Piarulli:2017dwd,Baroni:2018fdn}, all considered at the next-to-next-to-next-to-leading-order (N3LO) in the chiral expansion.  We use of five classes of them labeled as Ia, Ia$^*$, Ib$^*$, IIa$^*$, and IIb$^*$. 
Three choices for fitting the LECs give rise to the different classes of the Norfolk model. The roman numeral I and II indicates the energy range used to fit the two-nucleon interaction, and the letters a and b indicate the choice of local regulator functions. Details on these choices can be found in Ref.~\cite{Piarulli:2016vel}. Stars indicate three-nucleon interactions fit with strong-plus-weak interaction data~\cite{Baroni:2018fdn}, while no star indicates a fit with strong interaction data only~\cite{Piarulli:2017dwd}.

To compute the form factors, we use the charge  and current  operators derived 
in $\chi$EFT~\cite{Kolling:2009iq,Kolling:2011mt,Pastore:2008ui,Pastore:2009is,Pastore:2011ip,Gnech:2022vwr}, specifically those of Refs.~\cite{Pastore:2011ip,Gnech:2022vwr} which are consistent with the Norfolk interaction. They contain two-body currents that are known to be  important in electromagnetic observables~\cite{Miyagi:2023zvv,Chambers-Wall:2024fha,Chambers-Wall:2024uhq,King:2024jiq,sun2026abinitiochargeform}.  For the GFMC calculations, we adopt the Norfolk model Ia, which successfully reproduces the static properties of light nuclei~\cite{Piarulli:2017dwd}, and is therefore well-suited to describe low-energy properties probed in this study. Model uncertainties are estimated from the spread of VMC calculations across different interaction models, following Ref.~\cite{King:2021jdb}, and the same relative error is assumed as a conservative estimate for the GFMC results. Although this provides a rough idea of model dependence, it does not represent a full exploration of the parameter space. Ongoing efforts to construct order-by-order local chiral interactions~\cite{Bub:2024gyz,Somasundaram:2023sup} and emulate calculations~\cite{Somasundaram:2024zse,Armstrong:2025tza} could enable for more sophisticated uncertainty quantification in the future.

\begin{table*}
\centering
\small 
\setlength{\tabcolsep}{3pt} 
\renewcommand{\arraystretch}{0.9} 
\begin{tabular}{@{} l l l l l l l l @{}} \hline \hline
Nucleus & Method & $\avg{R_Z}$ & $R_E$ & $R_M$ & $\avg{R^3_E}_{(2)}$ & $\avg{R_E^4}$ & $\mu$ \\ 
 & & (fm) & (fm) & (fm) & (fm$^3$) & (fm$^4$) & (n.m.) \\ \hline
$^3$H  &VMC &$2.30(3)$ $[2.34(5)]$ &$1.76(4)$ $[1.73(4)]$ &$1.83(3)$ $[1.92(3)]$ & $20(1)$ $[19(1)]$ & $20(2)$ $[19(2)]$ &$2.97(2)$ $[2.59(1)]$ \\
&GFMC &$2.27(3)$ $[2.32(5)]$ &$1.72(4)$ $[1.69(4)]$ & $1.84(3)$ $[1.93(3)]$ &$19(1)$ $[18(1)]$ & $16(2)$ $[15(2)]$ & $2.94(2)$ $[2.53(1)]$ \\ 
\hline
& Exp  & -- & $1.755(86)$~\cite{Amroun:1994qj} & $1.840(181)$~\cite{Amroun:1994qj} & -- & -- & $2.97896$~\cite{Purcell:2010hka} \\ \hline \hline

$^3$He &VMC &$2.47(3)$ $[2.54(4)]$ &$1.94(3)$ $[1.92(3)]$ &$1.93(2)$ $[2.04(4)]$ &$28(1)$ $[28(2)]$ & $32(2)$ $[31(2)]$ &$-2.12(1)$ $[-1.77(-)]$\\
&GFMC &$2.50(3)$ $[2.57(4)]$ & $2.00(3)$ $[1.97(3)]$ & $1.93(2)$ $[2.04(4)]$ &$31(2)$ $[30(2)]$ &$36(3)$ $[35(3)]$ &$-2.12(1)$ $[-1.77(-)]$  \\  
\hline
& Exp  & $2.528(16)$~\cite{Sick:2014} & $1.9506(14)$~\cite{Shiner:1995zz} & $1.976(47)$~\cite{Amroun:1994qj} & $28.15(70)$~\cite{Sick:2014} & $32.9(1.6)$~\cite{Sick:2014} & $-2.12750$~\cite{Purcell:2010hka} \\ \hline \hline

$^4$He &VMC & -- & $1.68(2)$ $[1.65(2)]$   & -- &$17.1(9)$ $[16.3(9)]$  & $14.4(6)$ $[13.6(6)]$&-\\
      &GFMC & -- & $1.69(2)$ $[1.66(2)]$ & -- & $17.3(7)$ $[16.5(7)]$  & $14.4(6)$ $[13.7(6)]$ &- \\  
\hline
& Exp  &  -& $1.67824(83)$~\cite{Krauth:2021foz}  & - & $16.73(10)$~\cite{Sick:2014}& $14.35(11)$~\cite{Sick:2014}  &  -\\ \hline \hline

$^6$Li &VMC &$3.84(20)$ $[3.84(20)]$ &$2.58(9)$ $[2.56(9)]$ &$3.25(24)$ $[3.26(24)]$ &$67(11)$ $[66(10)]$ &$86(16)$ $[84(15)]$ &$-0.84(-)$ $[-0.82(1)]$ \\
&GFMC &$3.75(20)$ $[3.74(19)]$ & $2.61(9)$ $[2.59(9)]$ & $3.37(25)$ $[3.37(25)]$ & $68(12)$ $[67(11)]$ &$84(15)$ $[82(15)]$ &$-0.82(-)$ $[-0.80(1)]$ \\
\hline
& Exp  & $2.44(2)$~\cite{Sun:2023} & $2.589(39)$~\cite{Nortershauser:2011zz} & $3.12(22)$~\cite{DeJager:1974} & -- & -- & $-0.82205$~\cite{Tilley:2002vg} \\ 
&  & $2.40(16)$~\cite{Qi:2020} &  &  &  &  &  \\ 
&  & $2.47(8)$~\cite{Qi:2020} &  &  &  &  &  \\ 
&  & $2.30(3)$~\cite{Puchalski:2013} &  &  &  &  &  \\ \hline\hline

$^7$Li &VMC &$3.47(11)$ $[3.49(12)]$ &$2.44(7)$ $[2.42(7)]$ &$2.94(17)$ $[2.98(14)]$ &$55(6)$ $[54(6)]$ &$64(10)$ $[62(10)]$ &$3.32(4)$ $[2.92(3)]$ \\
&GFMC &$3.36(11)$ $[3.39(12)]$ &$2.38(7)$ $[2.36(7)]$ &$2.77(17)$ $[2.87(14)]$ &$51(6)$ $[49(5)]$ &$56(9)$ $[54(9)]$ & $3.19(4)$ $[2.79(4)]$ \\
\hline
& Exp  & $3.33(7)$~\cite{Qi:2020} & $2.444(42)$~\cite{Nortershauser:2011zz} & $2.80(8)$~\cite{DeJager:1974}  & -- & -- & $3.25644$~\cite{Tilley:2002vg} \\ 
&  & $3.38(3)$~\cite{Qi:2020} &  &  &  &  &  \\ 
&  & $3.25(3)$~\cite{Puchalski:2013} &  &  &  &  &  \\ \hline\hline

$^9$Be  &VMC &$3.64(39)$ $[3.69(44)]$ &$2.48(5)$ $[2.47(5)]$ &$3.21(19)$ $[3.34(19)]$ &$57(5)$ $[56(5)]$ &$64(4)$ $[62(4)]$ &$-1.03(16)$ $[-0.96(17)]$ \\
& GFMC &$3.98(43)$ $[3.89(46)]$ &$2.53(5)$ $[2.52(5)]$ &$3.73(22)$ $[3.57(20)]$ &$58(4)$ $[58(3)]$ &$57(4)$ $[58(4)]$ & $-1.22(19)$ $[-1.02(18)]$ \\
\hline
& Exp  & $4.07(5)(2)~\cite{Puchalski:2014}$ & $2.519(12)$~\cite{Jansen:1972iui} & $2.67(6)$~\cite{DeJager:1974} & -- & -- & $-1.1778(9)$~\cite{Tilley:2004zz} \\ 
&  & $4.048(3)$~\cite{Dickopf:2024} &  &  &  &  &  \\ \hline\hline
\end{tabular}
\caption{Comparison of VMC and GFMC results for the nuclei under study using the Ia model of the nuclear interaction to experimental data. Calculations performed using charge and current operators up to N3LO in the chiral expansion are shown next to bracketed values computed in the impulse approximation. Note that the wave functions in the calculation are always generated with the full N3LO nuclear interaction. Values in parentheses indicate either the combined numerical, statistical, and model uncertainty from the extraction of the Monte Carlo results, or experimental uncertainty for measurements. Statistical uncertainties $<0.01$ on the Monte Carlo calculations have been omitted.}
\label{tab:zemach.gfmc}
\end{table*}

\section{Results}
\label{sec:results}

\begin{figure*}[tbh]
	\centering
   \includegraphics[width=0.7\textwidth]{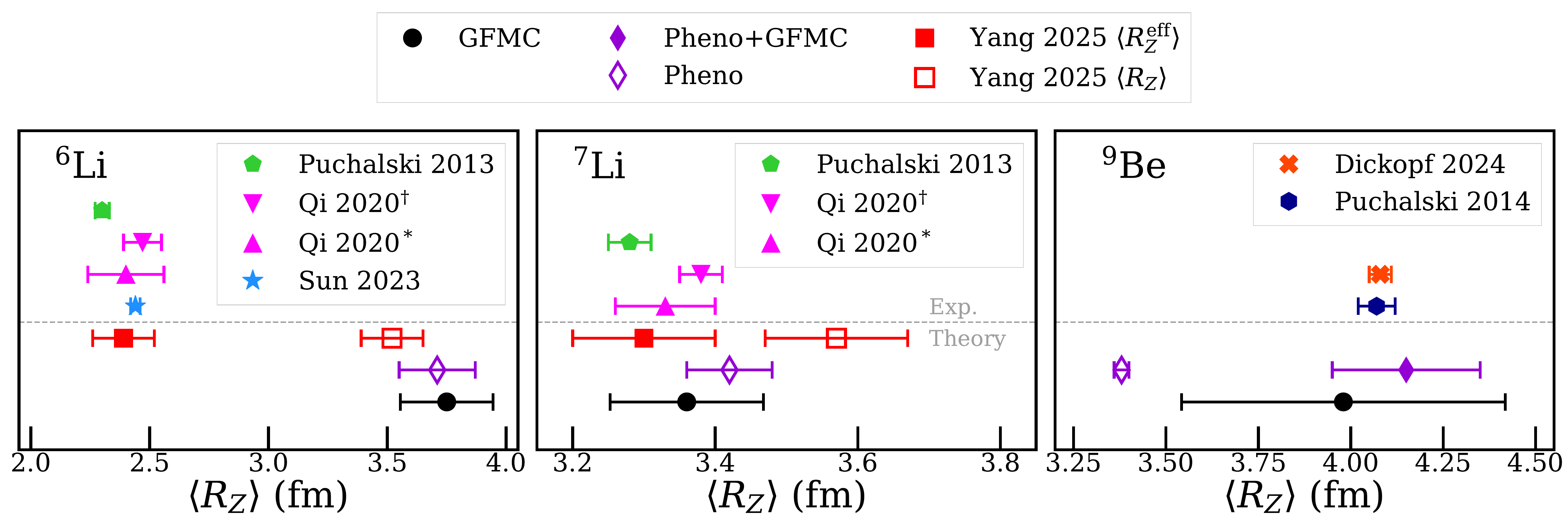}
	\caption{$\avg{R_Z}$ values from various theoretical approaches and experimental measurements for $^6$Li, $^7$Li, and $^9$Be. The filled black circles are the GFMC results obtained with N3LO charge and current operators for this interaction. The error bars on these calculations are estimated from the spread of the VMC calculations in this work using different nuclear models. The open purple diamonds represent the values using the phenomenological formula presented in Ref.~\cite{Yerokhin:2008} and experimental electromagnetic radii. The solid purple diamond indicates the phenomenological formula using the charge radius data and the magnetic radius from GFMC (denoted Pheno+GFMC). The open red squares represent $\avg{R_Z}$ from Ref.~\cite{Yang:2025rcx} computed with a VMC approach using a neural-network ansatz, while the filled red squares represent results from the same work including an estimate of inelastic two-photon exchange effects. Experimental measurements are represented with blue stars, magenta triangles, orange crosses, and green hexagons. The data are from Refs.~\cite{Sick:2014,Sun:2023,Qi:2020,Puchalski:2013,Puchalski:2014,Dickopf:2024} and correspond to the symbols as indicated in the key of each panel. For the measurements of Qi {\it el al.}, we use a dagger or star to indicate different hyperfine splittings; namely, for $^6$Li, $*$ = $2^3S_1^0$ - $2^3S_1^1$, $\dagger$ = $2^3S_1^1$ - $2^3S_1^2$ and, for $^7$Li, $*$ = $2^3S_1^{3/2}$ - $2^3S_1^{5/2}$, $\dagger$ = $2^3S_1^{1/2}$ - $2^3S_1^{3/2}$. \label{fig:rz.panels}} 
\end{figure*}

The VMC results for the Zemach and electromagnetic moments, provided in the supplementary material~\cite{supplement} for completeness, are computed with the five Norfolk interactions introduced in Sec.~\ref{sec:model}. These calculations serve to estimate a model dependence band: electromagnetic radii and magnetic moments agree at the percent level across nuclei and models, with variations in $R_M$ and $\avg{R_Z}$ reaching the few percent level for systems with $A\geq 6$, and up to 11\% and 12\% for the magnetic moment and $\avg{R_Z}$ of $^9$Be, respectively. We see more variation in the terms that depend on higher order moments of the distributions, such as $\avg{R_E^4}$. This is consistent with {\it ab initio} studies in heavier nuclei~\cite{Door:2024qqz,Miyagi:2025rvx}, and reflects the fact that the variational wave function, while reproducing the cluster structure in light nuclei and capturing long-range correlations~\cite{Carlson:2014vla}, is primarily sampled near the interior, leaving the behavior of the tail less well-constrained by the optimization. Precise determinations of this quantity from electron scattering could help to put further constraints on the nuclear models used to predict electromagnetic quantities~\cite{Hiyama:2024zfd}, and variational ans\"{a}tze that impose the long-range behavior of the wave function could help better constrain this quantity~\cite{Nollett:2000ch, Nollett:2001ub,Flores:2022foz,Flores:2025kcr}.  We also note that at the VMC level, $\avg{R_Z}$ is not brought into agreement with the experimental data for $^6$Li, and is much closer to the standard nuclear physics value of Table~\ref{tab:nucl.phys.val}. 

To refine the VMC results, we perform a GFMC propagation of the model Ia wave functions. We compare the VMC and GFMC results with the experimental data in Table~\ref{tab:zemach.gfmc}. The GFMC propagation leaves $R_E$ rather stable in all nuclei, with few percent variations seen, keeping good agreement with the experimentally measured values, and $\mu$ agrees at the few percent level,  consistent with Ref.~\cite{Chambers-Wall:2024uhq}. The moments of the $^4$He distributions all show good agreement with the experiment, while $R_M$ shows larger variations for the $A\geq 6$ nuclei. 

We turn now our attention to the Zemach radii. Across all nuclei, the GFMC propagation improves over VMC, moving the central values of $\avg{R_Z}$ closer to the experimental data. For the $A\geq 6$ nuclei studied in this work, we plot our GFMC  results for $\avg{R_Z}$ (solid black circles) for comparison with experimental data (blue stars, magenta triangles, orange crosses, and green hexagons), standard nuclear physics values~\cite{Yerokhin:2008} (open purple diamonds), and the recent neural network results from Ref.~\cite{Yang:2025rcx} (open and filled red squares) in Figure~\ref{fig:rz.panels}. We stress that the errors on the VMC and GFMC results reflect the spread of VMC calculations across Norfolk models; improvements in uncertainty quantification would be valuable to match the current experimental precision. 

The case of $^3$He provides a clean benchmark: the full GFMC results retaining two-body currents up to N3LO in the chiral expansion agrees with the electron scattering analysis of Ref.~\cite{Sick:2014}. Two-body currents reduce the leading-order (LO) GFMC result by ${\sim}2.7\%$. The computation of $\avg{R_Z}$ without inelastic two photon exchange~\cite{Yang:2025rcx} agrees within error with the full GFMC result, with about a ${\sim 1\%}$ variation in the central values; however, with the estimation of polarizability contributions (filled red square), the authors of Ref.~\cite{Yang:2025rcx} find a ${\sim 5\%}$ smaller effective $\avg{R_Z}$ than the GFMC result.
Polarizability effects involve the excitation of intermediate nuclear states, and can be computed in formalisms that take nuclear response functions as input~\cite{Ji:review}. Computing polarizability contributions using GFMC Euclidean responses~\cite{Lovato:2016gkq,Lovato:2017cux,Lovato:2020kba} would be useful to size their effect, but is beyond the scope of the present work. 

Two results are worth highlighting. First, the disagreement between nuclear theory and atomic spectroscopy for $^6$Li persists at the GFMC
level, with our calculations favoring the standard nuclear physics value, in agreement with the independent neural network VMC calculations of Ref.~\cite{Yang:2025rcx} (open red squares), which uses local chiral interactions at N2LO and neglects two-body currents, obtaining $\avg{R_Z}= 3.65(1)$fm and $3.39(2)$ fm for a regulator value of $1.0$ and $1.2$ fm, respectively~\cite{Pengwei:privcomm}--smaller than our results by 5-10 $\%$ percent, even in the impulse approximation, as can be appreciated in panel (a) of the figure. 
The convergence of these independent approaches suggests that the discrepancy with the experimental data is not an artifact of the theory. Notably, Ref.~\cite{Yang:2025rcx} shows that including two-photon exchange effects brings their results for $^6$Li (filled red squares) in agreement with the data, implying that polarizability may be essential to resolve the discrepancy. The case of $^7$Li shown in panel (b) supports the claim that polarizability contributions are larger in odd-odd nuclei~\cite{Yang:2025rcx}: our GFMC calculations retaining two-body currents--without two-photon exchange effects--agrees with experiment at the $\lesssim 1\%$ to ${\sim 3\%}$ level across the various experimental measurements, and at the ${\sim 2\%}$ level with the standard nuclear physics result, which suggests that polarizability effects are negligible in this odd-$A$ system. There is a ${\sim 6\%}$ difference between our $^7$Li N3LO GFMC result and that of Ref.~\cite{Yang:2025rcx} (3.67(2) fm and 3.47(1) fm, 
for a regulator value of 1.0 and 1.2 fm, respectively~\cite{Pengwei:privcomm}). The inclusion of the effective polarizability contribution (filled red squares), while smaller than for $^6$Li, provides better agreement with experiment using the approach of Ref.~\cite{Yang:2025rcx}. We note that two-body contributions have a $\lesssim 1\%$ effect on the $\avg{R_Z}$ value in $^6$Li and $^7$Li. 

Second, for $^9$Be shown in panel (c), the N3LO GFMC result agrees with the experimental measurements within the error bar, with central values within $\sim2\%$, while the standard nuclear physics value~\cite{DeJager:1974} is discrepant by $\sim17\%$. The origin of this discrepancy can be traced to the magnetic radius entering the formula of Ref.~\cite{Yerokhin:2008}. Specifically, the value extracted from electron scattering in Ref.~\cite{DeJager:1974} relied on a simplified shell model analysis and differs from our {\it ab initio} GFMC result by $\sim40\%$~\cite{King:2025akz,Chambers-Wall:2024fha}. If we adopt the magnetic radius from the {\it ab initio} GFMC calculation in the standard nuclear physics formula of Ref.~\cite{Yerokhin:2008}, we obtain $\avg{R_Z} = 4.15(20)$ fm (indicated in the figure by the solid purple diamond), which is only ${\sim 2\%}$ away from the experimental numbers. More precise determinations of the magnetic radius from nuclear magnetic resonance~\cite{Bissell:2026umt} and molecular spectroscopy~\cite{Wilkins:2025} would complement the electron scattering analysis, which requires modeling and therefore do not provide an unambiguous experimental value for this quantity. The good agreement of the GFMC result with the data again supports the conjecture of Ref.~\cite{Yang:2025rcx} that inelastic two-photon exchange effects should be small in odd-$A$ nuclei, vanishing in the limit of SU(4) symmetry, which could explain why the $\avg{R_Z}$ computed in this work agrees with the experimental extraction. Finally, we note that two-body currents account for a small enhancement of $\avg{R_Z}$ by approximately $2\%$. 

\begin{figure}[tbh]
	\centering
   \includegraphics[width=0.9\columnwidth]{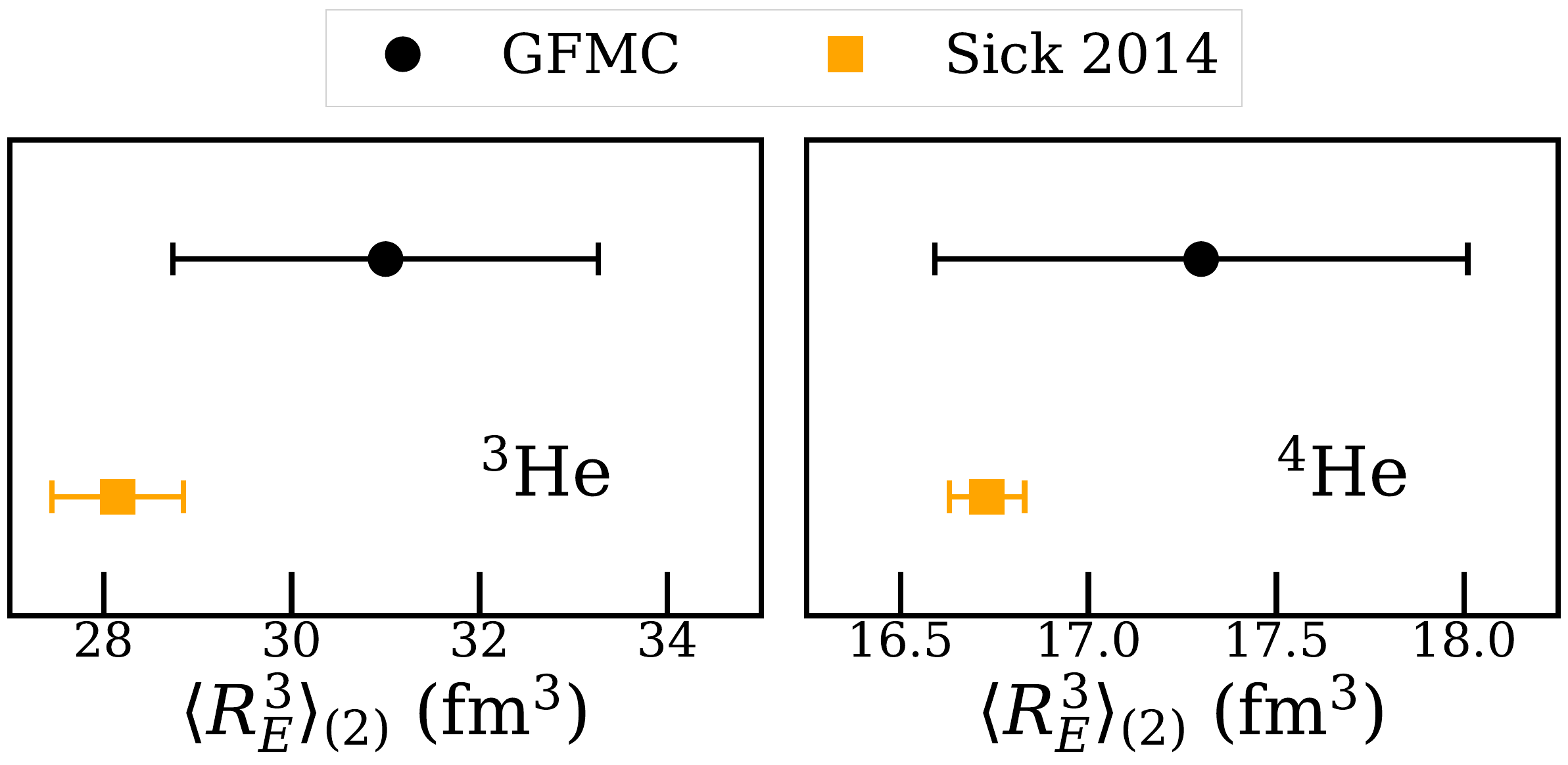}
	\caption{The moment $\avg{R^3_E}_{(2)}$ from GFMC with the N3LO charge operator (filled black circles) compared with values extracted from electron scattering measurements of the $^3$He and $^4$He charge form factors (orange squares). The theoretical error bars are from the estimated model uncertainties based on the analysis of VMC results in Table~\ref{tab:zemach} and the statistical uncertainty of the fit, while the experimental error bar comes from the random and systematic errors of the data which are discussed in the analysis of Ref.~\cite{Sick:2014}. \label{fig:re3}} 
\end{figure}

Finally, in Figure~\ref{fig:re3}, we present the LO and N3LO GFMC third electric Zemach moments $\avg{R^3_E}_{(2)}$ relative to the values extracted from the analysis of electron scattering for $^3$He and $^4$He performed in Ref.~\cite{Sick:2014}. Two-body currents have a ${\sim} 3\%$ and ${\sim} 5\%$ enhancement for $^3$He and $^4$He, respectively. The uncertainties shown in the Figure combine the statistical uncertainty of the fit with the estimate of the model uncertainty from the VMC calculations. Within the uncertainties, we have agreement with the extraction from experiment; however, the N3LO GFMC result disagrees with the central value of the extraction by ${\sim} 10\%$ for $^3$He and by ${\sim} 3.4\%$ for $^4$He. As discussed for the VMC calculations, the higher-order moments of the charge distribution that $\avg{R^3_E}_{(2)}$ depends on are more variable in our calculations. This is again likely due to the fact that the tail of the wave function is not sampled as much as the interior of the wave function in the Monte Carlo. Future studies of different trial wave functions for the GFMC that put more constraint on the long-range structure could help reduce variability in these moments. This will be particularly impactful for extractions of charge radii from the Lamb shift~\cite{Quartet}, which require nuclear structure input for two-photon exchange contributions. The third electric Zemach moment is tied to the elastic two-photon exchange contribution~\cite{Ji:review} and is an important ingredient in extracting these radii.

\section{Conclusions}

We presented new ab initio calculations of the Zemach radii of light nuclei, and compare them with other data in the literature. 
In case of the $^{6,7}$Li nuclei, we examine the origin of the discrepancy between atomic physics extractions and standard nuclear physics results. Our findings suggest that the disagreement may not stem from deficiencies in nuclear theory calculations, but rather hint towards issues in the experimental extraction methods themselves. This underscores the need for a careful reassessment of isotope shift analyses in lithium, akin to the reevaluations that resolved similar puzzles in other light nuclear systems. In $^9$Be, we highlighted that different models that can model the magnetic form factor can produce different radii. Since $R_M$ is not an unambiguously determined experimental quantity, having further input from other {\it ab initio} nuclear approaches would be of great benefit to atomic physics.

From the perspective of nuclear physics, the Zemach radius provides complementary information to the charge radius, probing both the spatial distribution of nuclear magnetization and its correlation with the charge density. It thus serves as a benchmark for nuclear models and ab initio calculations aiming to describe nuclear electromagnetic structure with controlled theoretical uncertainties. Explicit inclusion of two-body currents indicated that the effect is rather small on $\avg{R_Z}$ in light nuclei. It would be interesting to have similar computations in heavier systems using other many-body methods that can treat two-body contributions to see if the trend continues. 

Higher-order moments of the charge distribution, which are related to the determination of the third electric Zemach moment $\avg{R^3_E}_{(2)}$, are not as well-constrained across models in the VMC calculations. This is likely-due to the tail of the wave function being sampled less than the interior in the Monte Carlo process. In the future, designing trial wave functions that impose more of a constraint on the behavior of the tail will help to improve the stability of these results. This will be important for providing information on the elastic to two-photon exchange contributions to the Lamb shift, which is relevant for determining charge radii from precision spectroscopy. 

Overall, this work provides an {\it ab initio} quantification of elastic nuclear structure effects for hyperfine splittings and Lamb shifts. Because of recent advances in optical spectroscopy of atoms and ions, as well as in high-precision X-ray spectroscopy, reliable structure information is useful. Our structure calculations highlight possible sources of discrepancies in the analyses of the $^6$Li and $^9$Be hyperfine structure, demonstrating the usefulness of {\it ab initio} nuclear theory in this area. We have also highlighted future directions to further improve upon these calculations, which will support the potential for beyond the Standard Model physics searches using spectroscopy. 

\section*{Acknowledgements}
We acknowledge useful discussions with Chen Ji and thank Pengwei Zhao for useful comments on the manuscript. We thank Yilong Yang, Evgeny Epelbaum, Chen Ji, and Pengwei Zhao for sharing their neural network ansatz VMC data for use in Fig.~\ref{fig:rz.panels}. 
Financial support by Los Alamos
National Laboratory's Laboratory Directed Research and Development program under project 20240742PRD1 (G.~B.~K.)
is gratefully acknowledged. Los Alamos National Laboratory is operated by Triad National Security, LLC,
for the National Nuclear Security Administration of U.S.\ Department of Energy (Contract No.\
89233218CNA000001). 
We also acknowledge  support by the Deutsche Forschungsgemeinschaft (DFG) through the Cluster of Excellence ``Precision Physics, Fundamental Interactions, and Structure of Matter'' PRISMA${}^+$ EXC 2118/1 (Project ID 390831469) and through CRC1660: Hadrons and Nuclei as discovery tools (Project No. 514321794) (S.~B.).
This work is also supported by the US Department of Energy under Contracts No. DE-SC0021027 (S.P. and G.~C.~W.), DE-AC02-06CH11357 (R.B.W.), DE-AC05-06OR23177 (A.G.), a 2021 Early Career Award number DE-SC0022002 (M.~P.), the FRIB Theory Alliance award DE-SC0013617 (M.~P.), and the NUCLEI SciDAC program (S.P., M.P., and R.B.W.).  A.G. acknowledges the direct support of Nuclear Theory for New Physics Topical collaboration.

\bibliographystyle{elsarticle-num}
\bibliography{biblio}

\newpage
\begin{center}
\textbf{\large Supplementary Material}
\end{center}

In order to estimate the model uncertainty on the more computationally intensive GFMC calculations, we performed VMC calculations with several nuclear models introduced in the text; namely, the Norfolk models Ia, Ia*, Ib*, IIa*, and IIb*. In this supplemental material, we present the VMC results that were used to estimate the model uncertainty on our calculation in Table~\ref{tab:zemach}.

\begin{table*}
\centering
\resizebox{\textwidth}{!}{
\begin{tabular}{ l l c c c c c c } \hline \hline
Nucleus & Model & $\avg{R_Z}$ (fm) & $R_E$ (fm) & $R_M$ (fm) & $\avg{R^3_E}_{(2)}$ (fm$^3$) & $\avg{R_E^4}$ (fm$^4$) & $\mu$ (n.m.) \\ \hline

$^3$H  
& Ia   & $2.30$ $\{2.34\}$ & $1.76$ $[1.73]$ & $1.83$ $[1.92]$ & $20(1)$ $[19.4(7)]$ & $19.7$ $[18.9]$ & $2.97$ $[2.59]$ \\ 
& Ia*  & $2.30$ $[2.32]$ & $1.74$ $[1.72]$ & $1.85$ $[1.91]$ & $20(1)$ $[19.0(7)]$ & $18.8$ $[18.0]$ & $2.97$ $[2.59]$ \\ 
& Ib*  & $2.30$ $[2.33]$ & $1.75$ $[1.72]$ & $1.84$ $[1.91]$ & $20(1)$ $[19.2(7)]$ & $19.4$ $[18.5]$ & $2.98$ $[2.58]$ \\
& IIa* & $2.27$ $[2.30]$ & $1.73$ $[1.70]$ & $1.82$ $[1.89]$ & $19(1)$ $[18.5(6)]$ & $18.3$ $[17.5]$ & $2.97$ $[2.59]$ \\
& IIb* & $2.27$ $[2.29]$ & $1.72$ $[1.69]$ & $1.82$ $[1.88]$ & $19(1)$ $[18.1(6)]$ & $17.7$ $[16.9]$ & $2.99$ $[2.59]$ \\
\hline 
& Exp  & -- & $1.755(86)$~\cite{Amroun:1994qj} & $1.840(181)$~\cite{Amroun:1994qj} & -- & -- & $2.97896$~\cite{Purcell:2010hka} \\ \hline \hline

$^3$He 
& Ia   & $2.47$ $[2.54]$ & $1.94$ $[1.92]$ & $1.93$ $[2.04]$ & $28(1)$ $[28(2)]$ & $31.7$ $[30.8]$ & $-2.12$ $[-1.77]$ \\
& Ia*  & $2.46$ $[2.55]$ & $1.96$ $[1.94]$ & $1.91$ $[2.05]$ & $29(1)$ $[28(2)]$ & $33.0$ $[32.1]$ & $-2.12$ $[-1.77]$ \\ 
& Ib*  & $2.48$ $[2.55]$ & $1.96$ $[1.94]$ & $1.92$ $[2.04]$ & $29(1)$ $[28(2)]$ & $32.4$ $[31.4]$ & $-2.12$ $[-1.77]$ \\ 
& IIa* & $2.46$ $[2.53]$ & $1.94$ $[1.92]$ & $1.92$ $[2.03]$ & $28(1)$ $[28(2)]$ & $31.8$ $[30.9]$ & $-2.12$ $[-1.77]$ \\ 
& IIb* & $2.45$ $[2.51]$ & $1.93$ $[1.91]$ & $1.91$ $[2.01]$ & $28(1)$ $[27(1)]$ & $30.7$ $[29.8]$ & $-2.13$ $[-1.77]$ \\ 
\hline
& Exp  & $2.528(16)$~\cite{Sick:2014} & $1.9506(14)$~\cite{Shiner:1995zz} & $1.976(47)$~\cite{Amroun:1994qj} & $28.15(70)$~\cite{Sick:2014} & $32.9(1.6)$~\cite{Sick:2014} & $-2.12750$~\cite{Purcell:2010hka} \\ \hline \hline

$^4$He 
& Ia   &  -&  $1.68$ $[1.65]$ & - & $17.1(5)$ $[16.3(5)]$  & $14.4$ $[13.6]$ & -\\
& Ia*  &  -& $1.66$ $[1.63]$ & - & $16.6(5)$ $[15.8(4)]$  & $13.9$ $[13.1]$ &  -\\ 
& Ib*  &  -& $1.66$ $[1.63]$ & -  &  $16.7(5)$ $[15.8(4)]$& $14.0$ $[13.2]$ & - \\ 
& IIa* &  -&  $1.66$ $[1.63]$ & - & $16.4(5)$ $[15.6(4)]$  & $13.8$ $[13.0]$  & - \\ 
& IIb* &  -& $1.67$ $[1.64]$  & - & $16.9(5)$ $[16.0(4)]$ & $14.3$ $[13.4]$  & -\\ 
\hline
& Exp  &  -& $1.67824(83)$~\cite{Krauth:2021foz}  & - & $16.73(10)$~\cite{Sick:2014}& $14.35(11)$~\cite{Sick:2014}  &  -\\ \hline \hline

$^6$Li 
& Ia   & $3.84$ $[3.84]$ & $2.58$ $[2.56]$ & $3.25$ $[3.26]$ & $66(6)$ $[67(6)]$ & $84.1(1)$ $[86.0(1)]$ & $-0.84$ $[-0.82]$ \\
& Ia*  & $3.86$ $[3.86]$ & $2.58$ $[2.56]$ & $3.32$ $[3.32]$ & $67(6)$ $[66(6)]$ & $85.3(1)$ $[83.4(1)]$ & $-0.84$ $[-0.83]$ \\
& Ib*  & $3.96$ $[3.95]$ & $2.63$ $[2.61]$ & $3.39$ $[3.40]$ & $73(7)$ $[71(7)]$ & $96.9(1)$ $[94.9(1)]$ & $-0.84$ $[-0.82]$ \\ 
& IIa* & $3.76$ $[3.75]$ & $2.54$ $[2.52]$ & $3.15$ $[3.16]$ & $64(6)$ $[63(6)]$ & $81.4(1)$ $[79.6(1)]$ & $-0.84$ $[-0.82]$ \\
& IIb* & $3.88$ $[3.88]$ & $2.61$ $[2.59]$ & $3.29$ $[3.30]$ & $68(6)$ $[70(6)]$ & $88.7(1)$ $[86.7(1)]$ & $-0.84$ $[-0.82]$ \\ 
\hline
& Exp  & $2.44(2)$~\cite{Sun:2023} & $2.589(39)$~\cite{Nortershauser:2011zz} & 3.12(22)~\cite{DeJager:1974} & -- & -- & $-0.82205$~\cite{Tilley:2002vg} \\ 
&  & $2.40(16)$~\cite{Qi:2020} &  &  &  &  &  \\ 
&  & $2.47(8)$~\cite{Qi:2020} &  &  &  &  &  \\ 
&  & $2.30(3)$~\cite{Puchalski:2013} &  &  &  &  &  \\ \hline\hline
$^7$Li 
& Ia   & $3.47$ $[3.49]$ & $2.44$ $[2.42]$ & $2.94$ $[2.98]$ & $54(4)$ $[55(4)]$ & $64.1$ $[62.3]$ & $3.32$ $[2.92]$ \\
& Ia*  & $3.49$ $[3.51]$ & $2.45$ $[2.43]$ & $3.00$ $[3.02]$ & $56(4)$ $[55(4)]$ & $67.6$ $[65.7]$ & $3.31$ $[2.93]$ \\
& Ib*  & $3.39$ $[3.41]$ & $2.39$ $[2.37]$ & $2.89$ $[2.90]$ & $51(3)$ $[50(3)]$ & $57.7$ $[55.8]$ & $3.32$ $[2.91]$ \\
& IIa* & $3.43$ $[3.44]$ & $2.40$ $[2.38]$ & $2.87$ $[2.94]$ & $53(4)$ $[51(4)]$ & $60.0$ $[58.2]$ & $3.28$ $[2.90]$ \\
& IIb* & $3.38$ $[3.39]$ & $2.38$ $[2.36]$ & $2.83$ $[2.88]$ & $51(3)$ $[50(3)]$ & $57.9$ $[56.0]$ & $3.32$ $[2.93]$ \\ 
\hline
& Exp  & $3.33(7)$~\cite{Qi:2020} & $2.444(42)$~\cite{Nortershauser:2011zz} & $2.80(8)$~\cite{DeJager:1974} & -- & -- & $3.25644$~\cite{Tilley:2002vg} \\ 
&  & $3.38(3)$~\cite{Qi:2020} &  &  &  &  &  \\ 
&  & $3.25(3)$~\cite{Puchalski:2013} &  &  &  &  &  \\ \hline\hline
$^9$Be 
& Ia   & $3.64$ $[3.69]$ & $2.48$ $[2.47]$ & $3.21$ $[3.34]$ & $57(3)$ $[56(4)]$ & $63.6$ $[62.1]$ & $-1.03$ $[-0.96]$\\
& Ia*  & $3.34$ $[3.35]$ & $2.46$ $[2.44]$ & $3.29(7)$ $[3.44(2)]$ & $55(3)$ $[54(4)]$ & $61.4(1)$ $[59.4]$ & $-1.04$ $[-0.94]$\\
& Ib*  & $3.73$ $[3.79]$ & $2.51$ $[2.49]$ & $3.40$ $[3.41]$ & $59(3)$ $[57(4)]$ & $64.8$ $[63.0]$ & $-1.19$ $[-1.11]$ \\
& IIa* & $3.62$ $[3.68]$ & $2.47$ $[2.45]$ & $3.24$ $[3.25]$ & $56(3)$ $[55(4)]$ & $60.6$ $[59.0]$ & $-1.17$ $[-1.09]$ \\
& IIb* & $3.69$ $[3.74]$ & $2.48$ $[2.46]$ & $3.30$ $[3.36]$ & $57(3)$ $[55(4)]$ & $62.1(1)$ $[60.1]$ & $-1.13$ $[-1.06]$ \\ 
\hline
& Exp  & $4.07(5)(2)~\cite{Puchalski:2014}$ & $2.519(12)$~\cite{Jansen:1972iui} & $2.67(6)$~\cite{DeJager:1974} & -- & -- & $-1.1778(9)$~\cite{Tilley:2004zz} \\ 
&  & $4.048(3)$~\cite{Dickopf:2024} &  &  &  &  &  \\ 
\hline \hline
\end{tabular} }
\caption{VMC results for Zemach moment retaining currents to N3LO in the chiral expansion for various nuclei. The result in the impulse approximation is included in brackets for comparison. Values in parentheses indicate either numerical uncertainty from the integration or statistical uncertainty from the extraction for Monte Carlo results, or experimental uncertainty for measurements. Statistical uncertainties $<0.01$ on VMC calculations have been omitted.}
\label{tab:zemach}
\end{table*}

\end{document}